\documentclass[preprint,12pt]{elsarticle}

\usepackage{graphicx}
\usepackage{printlen}

\usepackage{graphicx}
\usepackage{dcolumn}
\usepackage{bm}
\usepackage{amsthm}
\usepackage{float}

\usepackage{lineno}
\usepackage{hyperref}
\usepackage{color}
\usepackage{amsmath}
\usepackage{amssymb}
\usepackage{epstopdf}
\usepackage[short]{optidef}
\usepackage{cleveref}
\usepackage{comment}

\journal{Physica A: Statistical Mechanics and its Applications}

\begin{document}

\begin{frontmatter}


\title{Modeling the Spread of Multiple Contagions on Multilayer Networks}



\author[1]{Petar Jovanovski }
\author[1]{Igor Tomovski}
\author[1,2]{Ljupco Kocarev\corref{cor1} }
\cortext[cor1]{Corresponding author: Macedonian Academy of Sciences and Arts, P.O. Box 428, 1000 Skopje, Republic of Macedonia}

\address[1]{Macedonian Academy of Sciences and Arts, Republic of Macedonia}%
\address[2]{Ss. Cyril and Methodius University, Republic of Macedonia\\}%

\ead{lkocarev@manu.edu.mk}

\begin{abstract}
A susceptible-infected-susceptible (SIS) model of multiple contagions on multilayer networks is developed to incorporate different spreading channels and disease mutations. 
Upper/lower-bound basic reproductive numbers are derived and rapid mutation processes is analytically analyzed. In the special case when considering only two contagions, we analytically analyze models on multilayer networks with arbitrary number of layers, competition models on multilayer networks with arbitrary number of layers, and compartmental models. The novel multiple-contagion SIS model on a multilayer network could help in the understanding of other spreading phenomena including communicable diseases, cultural characteristics, addictions, or information spread through e-mail messages, web blogs, and computer networks.
\end{abstract}


\end{frontmatter}

\section{Introduction}
Epidemiological models, developed as tools for analyzing the spread and control of infectious diseases, have also been adapted to model the dynamics of contagious entities as diverse as communicable diseases, cultural characteristics (such as religious beliefs, fads or innovations), addictions, or information spread (through rumors, e-mail messages, web blogs, peer-to-peer computer networks, etc). The susceptible-infected-susceptible (SIS) model is one of the simplest and well-studied model for emerging disease outbreaks (like influenza, chlamydia, gonhorrea, etc.) that does not give immunity upon recovery \cite{Pastor2001,Dorogovtsev2008,Pastor2015,Mata2014,Chao2016}. The SIS compartmental model divides the population into two compartments (classes)  -- susceptible to the infection of the pathogen (often denoted by S) and infected by the pathogen (given the symbol I). The disease is transmitted with a rate $\beta$ while infected individuals become susceptible again with a rate $\gamma$.
The dynamics of the infectious class depends on a quantity called \textit{basic reproduction number}, $R_0$, defined as the number of secondary infections caused by a single infective introduced into a population made up entirely of susceptible individuals over the course of the infection of this single infective.  For the SIS model in a well-mixed population, the basic reproduction number is equal to $R_0 = {\beta}/{\gamma}$ \cite{Hethcote2000}.
By adopting so called quenched mean-field theory \cite{Piet2009,Arenas2013}, the basic reproduction number for the SIS model on networks has been estimated to be equal to  
\begin{equation} \label{eq-rep-number} 
R_0 =  \frac{\lambda_{\text{max}}(A) \beta}{\gamma} \equiv \frac{\beta_{\rm eff}}{\gamma},  
\end{equation}
where $A$ is the adjacency matrix of the network and $\lambda_{\rm max}(A)$ is the largest eigenvalue of the matrix $A$. The network structure, described with (expressed through) the largest eigenvalue of the network adjacency matrix, is encoded in the \textit{effective transmission rate} $\beta_{\rm eff} = \lambda_{\text{max}}(A) \beta$.    
For the SIS model in homogeneous networks, such as the Erdos-Renyi random networks and random regular graphs, 
$\lambda_{\rm max}(A) = \langle k \rangle $. According to Ref. \cite{Fan2003}, for scale-free networks,
$1/{\lambda_{\rm max}} = 1/\sqrt{k_{\rm max}}$ if $\hat{\gamma} > 5/2$, and  
$1/{\lambda_{\rm max}} = \langle k \rangle / \langle k^2 \rangle$ if $2 < \hat{\gamma} < 5/2$,  where the degree distribution follows $P(k)\sim k^{-\hat{\gamma}}$

In the thermodynamic limit when the number of nodes approaches infinity, a phase-transition occurs at $R_0=1$. If the basic reproduction number falls below the critical value ($R_0 < 1$), the infection dies out. For $R_0 > 1$ there is an epidemic in the population. At the phase transition, threshold for the transmission rate $\beta$ equals  
\begin{equation} \label{eq-beta-cr}
\beta_{cr} = \frac{\gamma}{\lambda_{\text{max}}(A) }. 
\end{equation}
Note that in Eq.~(\ref{eq-beta-cr}) for a fixed value of $\gamma$, large value of $\lambda_{max}(A)$ implies vanishing threshold ($\beta_{cr} \to 0$). In other words, the network structure causes an absence of an epidemic threshold and its associated critical behavior.

Models for multiple diseases that co-evolve in a network has recently been gaining attention \cite{Mark2005,Wei2013,Angel2014,Sahneh2014}. However, these diseases are often assumed to be mutually exclusive. While such models are usually discussed in the context of epidemics, they are more aptly used in studying belief propagation or product adoption, for example, in modeling competition in politics or competition in a marketplace.  The generalization of the SIS model to arbitrary number of multiple contagions, however, has not yet been developed. In this paper we model the spread of multiple contagions on networks.  We further assume that each contagion spreads over different spreading channels resulting in a \textit{multiple-contagion SIS model on a multilayer network}.

\subsection{Literature overview}

\subsubsection{Multiple-contagion compartmental models}

Developing models for interacting strains of the same pathogen, such as influenza \cite{Ferguson2003} or dengue \citep{Ferguson1999}, or interacting diseases such as HIV/AIDS and malaria \citep{Kublin2006}, is one of the most theoretically challenging problems in infectious disease epidemiology.  The central problem comes from the explosive growth in the number of state variables of the system with the linear increase in the number of strains or pathogens \citep{Gog2002}. 

The spread of multiple diseases can be characterized by various factors such as contact with infected individuals, the interplay between diseases coupled by mutation \cite{Abu2004} or cross-immunity \cite{Simone2009, Omori2010, Ferguson2002, Abu2004}, and coexistence \cite{Omori2010} or by the principle of competitive exclusion \cite{Bremermann1989, Carlos} which states that the strain with the largest reproduction number drives other strains into extinction. Epidemiological models that study multiple strains fall into super-infection models \cite{Alizon2013, Susi2015, NowakSuper} where strains cannot coexist in a host because the most virulent strain takes over, and co-infection models \cite{MayNowak1995, Susi2015} where coexistence is possible. For a discussion on how multiple infections have been modeled in evolutionary epidemiology, as well as co-and super-infection models in the standard setting, the reader is referred to \cite{Alizon2013}.  An epi-evolutionary model on a $n$-regular network with $N$ pathogens which are allowed to mutate into each other is proposed in \cite{SLion}. Besides spreading dynamics, the model takes into account the evolutionary properties of a parasite by keeping track of the change of the mean value. Authors demonstrate that the coupled dynamics lead to transient phenomena that cannot be described by standard invasion analyses.

Cross-immunity can be defined in at least two ways: one is reduced susceptibility and the other is reduced transmission (less likely to transmit) \cite{Kucharski2016}. The biological basis of this is that multiple influenza infections occur during the life of individuals and one's increased/decreased susceptibility depends on this history. The dimensionality of a history-based system is $O(2^N)$ (the power set of the pathogens) and multiple methods have been proposed for its reduction, focusing on a symmetrized system \cite{Abu2005, Ferguson2002} or via age-structure by grouping the population into compartments which have seen a strain $i$ \cite{Gog, StateSpace, Blyuss}. The work of \cite{Bichara} focuses on the stability analysis of a model of $n-$compartments, each having its own reproductive ratio and endemic threshold. Coexistence has been inspected in \cite{CoupledMS2016} when the strains are coupled by mutation. 
\subsubsection{Spreading processes on multilayer networks}
Multilayer networks are fundamental for the understanding of dynamical processes on networked systems, including, for example, spreading processes, such as flows (and congestion) in transportation networks \cite{Moris2012,Sole2016}, and information and disease spreading in social networks \cite{Wang2015,Funk2015,Granell2013,Sanz2014,Lima2015}.

As reviewed in \cite{Domenico2016}, there are two different categories of dynamical processes on multilayer networks: a single dynamical process and mixed (or coupled) dynamics, in which two or more dynamical processes, defined on each layer separately, are coupled together with inter-layer connections between nodes.
One of the simplest types of dynamics is a diffusion process. In a random walk, a discrete diffusion process, a walker jumps between nodes. In a multilayer network, the walker switches between layers via an inter-layer edge, resulting in enriched random-walk dynamics \cite{Mucha2010,Arenas2014,Radicchio2014}. The continuous diffusion process has also been analyzed in multiplex networks~\cite{Gomez2013,Sole2013} and novel phenomenon has been observed: diffusion can be faster in a multiplex network than in any of the layers considered independently.
Congestion in multilayer networks has been studied recently~\cite{Tan2014,Sole2016} for modeling multimodal transportation systems.

Coupled spreading processes on multilayer networks have recently been analyzed, including spreading dynamics of two concurrent diseases in two-layer multiplex networks~\cite{Dickison2012,Cozzo2013,Sanz2014,Salehi2015, jiang2018resource, jiang2018influence} and spread of disease coupled with the spread of information or behavior~\cite{Funk2009,Granell2013,Granell2014,Wang2015,Funk2015}. 
It was observed that two spreading processes can enhance each other (for example, one disease facilitates infection by the other), or one process can inhibit the spread of the other (for example, a disease can inhibit infection by another disease or the spreading of awareness about a disease can inhibit the spread of the disease). 

\subsection{Motivation and our contribution}
\begin{figure}
\begin{center}
\includegraphics[scale=0.75]{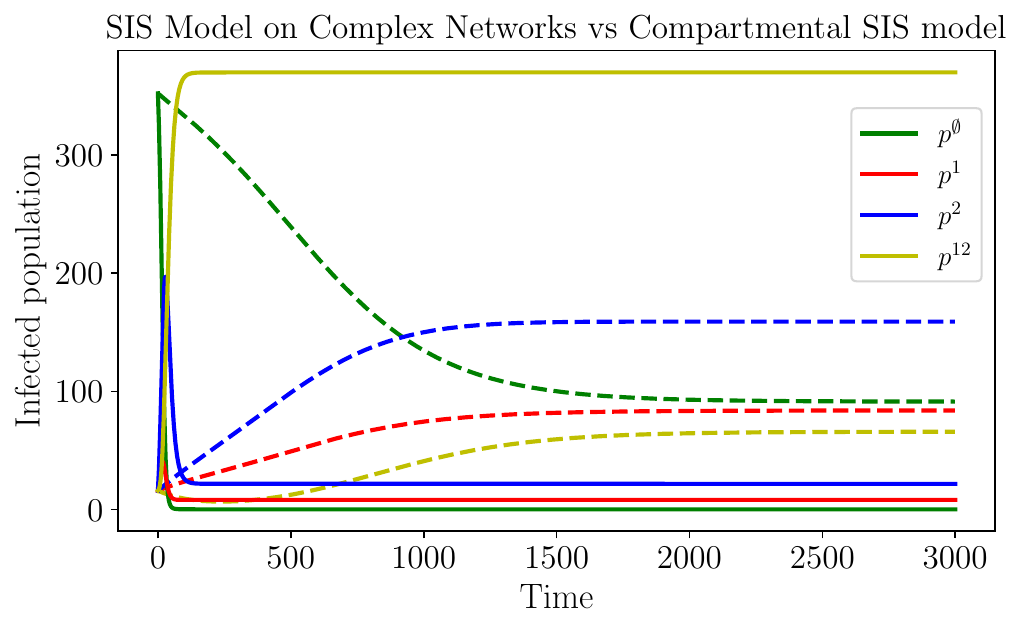}
\caption{The network over which the SIS network model is simulated is a two-layer Erdos-Renyi $G(n, p)$ network with $n = 400$ nodes and connection probability $p = 0.06$. The mutation $(\mu)$, transmission $(\beta)$ and recovery $(\gamma)$ rates are: $\mu_{12} = 0.01$, $\mu_{21} = 0.02$, $\beta^1_1= 0.02$, $\beta^1_2=0.01$, $\beta^2_1=0.02$, $\beta^2_2=0.04$, $\gamma_1 = 0.03$, $\gamma_2 = 0.01$. For a summarization of the preceding notation the reader is referred to table (\ref{table-1}). The solid lines in the graph present the infected subpopulation of the SIS model on the network, whereas the dashed lines present the subpopulation of the compartmental SIS model. Yellow presents the subpopulation infected with both (1 and 2) diseases, red and blue are the subpopulation infected with disease 1 and 2, respectively, and green is the susceptible population.}
\label{fig-netvscomp}
\end{center}
\end{figure}
This paper addresses multiple-contagion SIS model on multilayer networks.  So far only multiple-contagion compartmental models have been studied as reviewed in the previous subsection. However, a simple example illustrates that the SIS model on a complex network produces different dynamics from the compartmental SIS model with the same parameter values. Neglecting the network structure of the model and working only with the compartment model can lead to completely different results concerning how the infected populations evolve in time. Figure \ref{fig-netvscomp} depicts the infected sub-populations over time for two models: two-contagion compartmental SIS model and a two-contagion SIS model on a two-layer Erdos-Renyi random network. The number of nodes in the network model is equal to the total population in the compartmental model and both models have same parameter values. The equations governing the evolution of both the compartmental and network models are given in the next sections -- this is just an illustration that both models can exhibit different behavior and understanding multiple-contagion models on multilayer networks is the next step in modeling various spreading phenomena. 
The main contribution of this work is twofold: (1) a multiple-contagion SIS model on multilayer networks is developed; and (2) several properties of the model are analytically discussed including upper/lower bounds of basic reproduction number, rapid mutation processes and competition.
Furthermore, we discuss how the epidemic threshold depends on the network structure. In particular, two classes of the model are studied in detail. In the first class, contagions do not compete and a node can be infected by an arbitrary number of strains, while the second class describes competing epidemics on networks implying that each node (agent) can only be infected by a single contagion. Coinfection with multiple pathogens is a common occurrence. Some known combinations are HIV and tuberculosis, HIV and hepatitis, HIV and malaria. To the best of our knowledge, an epidemic threshold for two-contagions models and an arbitrary number of network layers has not yet been derived.
In a special case, we analytically analyze an example of a model with a mutation driven strain persistence characterized by the absence of an epidemic threshold.

\section{Model description}

We model multiple contagions on multilayer networks with three graphs: a multiplex graph, a bipartite graph, and a directed graph. The multiplex graph represents a social network consisting of a set of social actors and sets of dyadic ties.  A \textit{multiplex graph/network} is defined as a collection of a set of nodes $V$, a set $ R = \{1, \ldots, r \}$ of relation types, and for each $l \in { R}$, a set $E_l$ of edges describing the presence or absence of edges of type $l$ between pairs of nodes. The graph $(V,E_l)$ is also called a layer; we write $A_l = (A_{ij}^l)$ for the adjacency matrix of this graph. We assume that each layer represents a different spreading channel through which contagions are spread. For example, in etiology, scientists have recognized five major modes of disease transmission: airborne, waterborne, bloodborne, by direct contact, and through vector (insects or other creatures that carry germs from one species to another). These modes of disease transmission not necessarily spread over the same ties between individuals. Let the set $ H = \{1, 2 \ldots, m \}$ label the $m$ strains present in the system. For the bipartite graph $G= (R \cup H, E)$, the vertex set $R$ is the set of layers/channels, the vertex set $H$ is the set of contagions, and the set $E$ is the set of edges such that every edge connects a vertex in $R$ to one in $H$.  In a classical SIS model, the parameter $\beta$ represents contact or infection rate of a disease. Here $\beta^c_l$ denotes the contact or infection rate of the contagion/disease $c$ through the layer/channel $l$.  The bipartite graph can be represented with a weighted adjacency matrix defined by $[\beta^c_l]$.  We assume that contagions can be changed/mutated and be transformed between each other. For contagious diseases (also called communicable diseases), scientists have documented that diseases could mutate to become more contagious.    This is modeled as a directed graph with an $m \times m$ adjacency matrix $ [ \mu_{c b} ]$;  $ \mu_{cb}$ is a rate at which $c$ mutates into $b$.  In general, we assume that $\mu_{c b} \neq \mu_{b c}$. 

\begin{center}
\begin{table} 
\begin{center}
\begin{tabular}{ | l l |}
  \hline
  $R$ & Set of relation types.  \\
  $E_l$ & Edge set of layer $l$. \\
  $A_l$ & Adjacency matrix of layer $l$. \\
  $H$ & Set of contagions. \\ 
  $\beta^c_l$ & Infection rate of contagion $c$ on layer $l$. \\
  $\mu_{c b}$ & Mutation rate of contagion $c$ into contagion $b$. \\
  $\gamma_c$ & Healing rate of contagion $c$. \\
  $p^L_{i}(t)$ & Probability node $i$ is infected by all contagions \\
  & in $L$ at time $t$. \\ 
  $p^L = \sum_{i=1}^N p^L_{i}$ & Population infected by $L$. \\
  $L^c_b$ & Set $L$ including $c$, excluding $b$: $L \setminus b \cup c$. \\
  $R^c_0$ & Basic reproduction number for contagion $c$ \\
  $\beta^c_{\text cr}$ & Critical value for $\beta$ of contagion $c$. \\
  \hline
\end{tabular}
\end{center}
\caption{Summary of notation that is used throughout the paper.}
\label{table-1}
\end{table}
\end{center}

In the classical SIS model, when a single contagion is presented in the system, the node's state is defined with two variables: the probability that the node $i$ is susceptible $p^\emptyset_{i}$ and the probability that the node $i$ is infected $p^I_{ i}$. Clearly, these two variables are not independent since $p^\emptyset_{i} + p^I_{i} = 1$.  Here, for $m$ co-circulating contagions, the number of variables each node has is $2^m$, out of which $2^m -1$ are independent. The power set of ${ H}$, ${\cal P}({ H})$, is the set of all subsets of ${ H}$, including the empty set and ${ H}$ itself. Let $p^L_{i}$ be the probability that the node $i$ is currently infected (contaminated) by all contagious entities in the set ${ L}$, ${ L} \in {\cal P}({ H})$. Clearly, $\sum_{{ L} \in {\cal P}({ H} )} p^L_{i} = 1$.  The current contagious set of a node can be changed by one of three mechanisms. The first mechanism is mutation: a contagion mutates into another contagion. We assume that this mechanism is described with a rate: $\mu_{c b }$ denotes the rate at which $c$ mutates into $b$. The second mechanism is analogous to the node healing from contagion $c$ and $\gamma^c$ represents the recovery rate.  The third mechanism is transmission and is analogous to a node being infected with a disease from its neighbors. This mechanism is the only network-induced mechanism resulting in the change of the set ${ L}$ due to contact with the neighbors and we also refer to it as the contact-induced change mechanism of the set ${ L}$. This mechanism is described with a quantity $\beta^c_{l}$. Table \ref{table-1} summarizes all quantities used in the paper.

For $i = 1, \ldots, n$, $L \in {\cal P}({ H})$, and $|H| > |L| > 1$) the model consists of three parts. Equation (\ref{null}) captures the dynamics of node $i$ when it is infected by only one contagion (denoted by c). The first term describes the rate at which node $i$ recovers from contagion $l \neq c$ if it is infected with both $c$ and $l$. The second term describes the rate at which contagion $l$ mutates into contagion $c$ if node $i$ is infected by contagion $l$. The third term describes the rate at which node $i$ will be infected by contagion $c$ if it is susceptible. The fourth and fifth term describe the rate at which node $i$ will either recover from $c$ or $c$ will mutate into $l \neq c$, if it is infected with $c$. The fifth term describes the rate at which node $i$ will be infected by another contagion $l$, if it is already infected with $c$. 
\begin{equation} 
\dot{p}^c_{i} = \sum_{l \in H_c} p^{cl}_i \gamma_l +  p^l_i \mu_{lc} + \left(1 - \sum_{l \in \mathcal{P}(H) \setminus \emptyset} p^l_i \right) f^c_i - p^c_i \left(\gamma_c + \sum_{l \in H} \mu_{cl} + f^l_i \right), 
\label{null}
\end{equation}
where $f^c_i$ and $f^l_i$ are quantities defined with (\ref{transmission}).     
Equation (\ref{L}) has a more complex mechanism. The first term describes the rate at which node $i$ recovers from $c$ if it is infected by all contagions in $L$ and also $c$. The second term describes the rate at which contagion $c \in H_L$ will mutate into contagion $b \in L$, if node $i$ is infected with contagions $L \cup c \setminus b$. The third term describes the rate at which node $i$ will be infected by contagion $c$ if it is already infected with contagions $L \setminus c$. The fourth term is the rate at which node $i$ will recover from $c$ if it is infected by all contagions in $L$. The fifth term is the rate at which node $i$ will be infected by contagion $c \in H_L$ if it is infected will all contagions in $L$. The sixth term describes the rate at which contagion $c \in L$ will mutate into contagion $b \in L$. 
\begin{equation} 
\dot{p}^L_{i} =
\sum_{c \in H_L} \left( p^{L^c}_{i}\gamma_c + \sum_{b \in L}  p^{L^c_b}_{i} \mu_{c b} \right)
+ 
\sum_{c \in L} p^{L_c}_{i}f^c_i - p^L_{i}\gamma_c
-  
p^L_{i}  \sum_{c \in H_L} \left( f^c_i + \sum_{b \in L} \mu_{bc} \right).   
%
\label{L}
\end{equation}
Equation (\ref{H}) describes the case when the node is infected by all contagions. The first term is the rate at which node $i$ will become infected by $c$ if it is infected by all contagions in $H_c$. The second term describes the rate at which node $i$ will recover from one contagions, if it is infected by all contagions. 
\begin{equation} 
\dot{p}^H_{i} =
\sum_{c \in H} p^{H_c}_{i} f^c_i - 
p^H_{i} \sum_{c \in H} \gamma_c.
\label{H}
\end{equation}
The contact mechanism defined in equation (\ref{transmission}) goes over all neighbors of node $i$ in all layers. Contagion $c$ will be transmitted with corresponding rate $\beta^{c}_l$ for a layer/channel $l$ for those neighbors of $i$ which are infected by $c$, that is $c \in A$ where $A \in P(H)$ is the contagion set.
\begin{align} 
f^c_i =  
\sum_{\substack{j = 1}}^n \sum_{l \in R} \sum_{\substack{A \in P(H) \\ [c \in A]}} A^{l}_{ij}  \beta^c_{l} p^A_{ j}.
\label{transmission}
\end{align}
\subsection{Simulation on Networks}
\begin{figure}
\begin{center}
\includegraphics[scale=0.75]{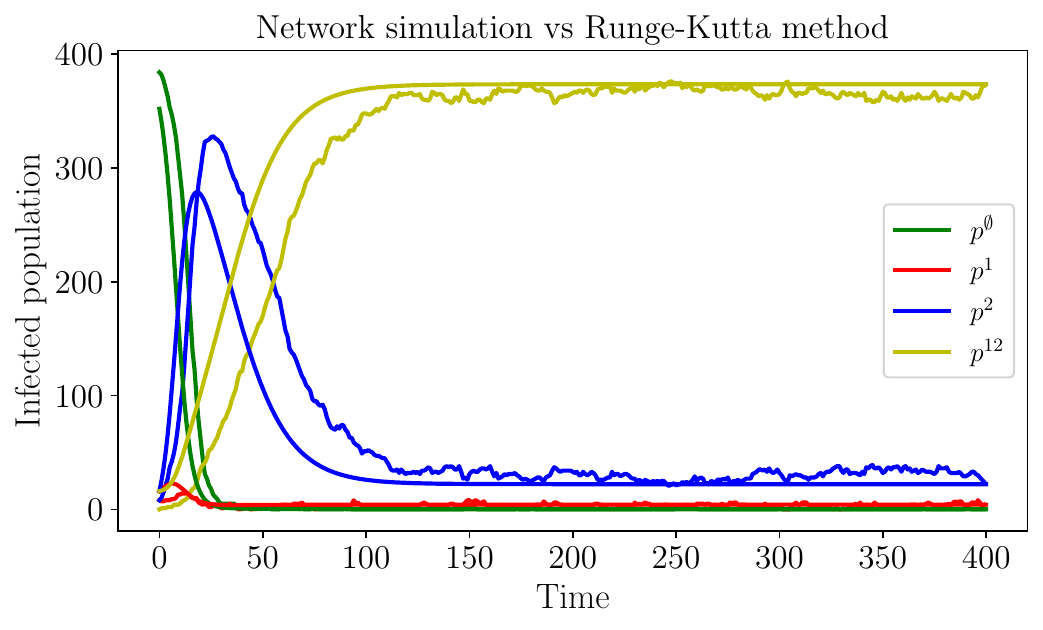}
\end{center}
\caption{Infected population versus time for deterministic and stochastic (wiggly line) two-contagion SIS model on a two-layer Erdos-Renyi random network. The deterministic model is described with equations (\ref{null}) -- (\ref{H}). The stochastic model is a discrete-time Markov process defined as follows. At time $t$ an infected node with contagion $c$ on layer $l$ can: (1) infect a susceptible neighboring node with rate $\beta^c_l \Delta t$, (2)  heal with rate $\gamma_c \Delta t$, and (3)  mutate with rate $\mu_{c b} \Delta t$. }
\label{SimVsRK}
\end{figure}

In this subsection we describe the Monte Carlo simulation procedure of the multiple contagion SIS model on multiple networks. The simulation is initialized by setting a certain percentage of nodes as infected. Then, for $t = 1, \ldots, T$, every infected (assume infected with all contagions in the set $L \subseteq H$) node tries to infect their neighbors with every contagion $c \in L$, via every channel. This is done by checking whether the neighbor is not infected with contagion $c$. If they aren't, the contagion is passed to the neighbor with probability $\Delta t \beta^c_l$. After this contact mechanism is finished, every infected node can either recover from contagion $c \in L$ with probability $\Delta t \gamma_c$, contagion $c \in L$ can mutate into contagion $d \in H \setminus L$ with probability $\Delta t \mu_{cd}$, or nothing occurs, and the node stays infected with the contagions in $L$. The corresponding continuous-time equations can be derived from this model leading to equations (\ref{null}) -- (\ref{H}). This derivation is, however, out of scope of the paper; a more detailed comparison of deterministic and stochastic models will be provided in a forthcoming paper. Figure \ref{SimVsRK} shows a comparison between the aforementioned Monte Carlo simulation and the fourth order Runge-Kutta method applied to equations (\ref{null}) -- (\ref{H}) with $\Delta t = 0.01$. The comparison is carried out on a two-contagion SIS model on two Erdos-Renyi random networks. A node in the network can be in either a susceptible state (S) or in one of the three infected states denoted as 1, 2, or 12 (12 means the node is infected both with contagions 1 and 2). 

\section{Properties of the model}
 
The model is a generalization of the classical SIS model on a single layer network. The equations (\ref{null})-(\ref{H})  assume that the population remains constant, i.e., the concentrations satisfy the closure condition $\sum_{L}p^L_{i}(t) = 1$ for every $t$ and every $i$. The equilibrium is obtained  by setting $\dot{p}^L_i = 0$. By acting so, the straightforward conclusion is that the point $p^L_{i}=0$, $L \in  P (H) \backslash {\emptyset}$, $p^L_{i}=1$ $i=1,..,n$, i.e the point of epidemic origin, is an equilibrium point for the dynamical system (\ref{null}), (\ref{L}), (\ref{H}), and (\ref{transmission}). 

It is a standard approach in the analysis of spreading processes occurring on complex networks, to derive the epidemic threshold relation, from the stability analysis of the point of epidemic origin. Locally stable epidemic origin implies that all initial perturbations of the system around the equilibrium point (epidemic outbreak) tend towards the origin (epidemic extinction) as time elapses. In the contrary, an unstable equilibrium point suggests that the outbreak will fully develop into persistent epidemic.

Due to the closure condition in the vicinity of the equilibrium, we restrict the stability analysis on the system of equations (\ref{L}). By performing a classical perturbation analysis around the point of epidemic origin, one obtains:
\begin{align} 
\frac{d\epsilon^c_{i}}{dt} &=
\sum_{l \in H} \epsilon^l_{i}  \mu^{l c} + \sum_{j=1}^n \epsilon^c_{j}\sum_{l \in R} A^l_{ij} \beta^c_l  -  \epsilon^c_{i} \left( \gamma_c + \sum_{l \in H } \mu_{c l} \right)\label{epsilonapp}\\
&=\sum_{l \in H} \epsilon^l_{i}  \mu_{l c} + \sum_{j=1}^n \epsilon^c_{j}t_{i,j}^c  -  \epsilon^c_{i}\left( \gamma_c + \sum_{l \in H } \mu_{c l} \right)\nonumber
\end{align}
where
\begin{equation}
t_{i,j}^c=\sum_{l \in R} A^l_{ij} \beta^c_l\label{tcij}
\end{equation}
The variables $\epsilon^c_{i}$ represent the probabilities that the node $i$ is infected by the strain $c$ at given instance in time, near the epidemic origin. One should note that the behavior of the system (\ref{null}), (\ref{L}), (\ref{H}), and (\ref{transmission})  near the epidemic origin restricts to the behaviour of a variant system in which only a single strain infection at each node $i$ may be present at a given time $t$. 

This simplifies the stability analysis of the system of equations (\ref{L}) from a system of dimension $n(2^m-1)$ to dimension $nm$.

Consider the vector $\mathbf{\epsilon}=[\epsilon^i_{j}]^T$. Let $\mathbf{D}=[d_{q,r}]$ be a $nm \times nm$ square matrix, which elements $d_{q,r}$ satisfy the following conditions:
\begin{itemize} 
\item $d_{q,r}=t_{i,j}^c$, if $q=cn+i$ and $r=cn+j$, $i \ne j$
\item $d_{q,r}=\mu_{l c}$, if $q=cn+i$ and $r=ln+i$ 
\item $d_{q,r}=-\gamma^c - \left(\sum_{l \in H } \mu_{c l} \right)$,  if $q=r=cn+i$
\item 0, otherwise.
\end{itemize}
Under these conditions, the system of equations (\ref{epsilonapp}), may be written in the following vector form:
\begin{equation} 
    \frac{d\mathbf{\epsilon}}{dt} = \mathbf{D}\mathbf{\epsilon} \label{epsvector}
\end{equation}
Before we proceed, notice that the matrix $\mathbf{D}$ is characterized by negative diagonal elements, while all off-diagonal entries are non-negative. For completeness, we will state and proof the following statement: Matrix $\mathbf{D}$ has a distinct eigenvalue $\lambda_1(\mathbf{D})$, such that:
\begin{itemize}
    \item $\lambda_1(\mathbf{D})$ is real;
    \item $\Re\{\lambda_j(\mathbf{D})\} < \lambda_1(\mathbf{D})$, $j=\overline{2,nm}$
\end{itemize}

We will refer to $\lambda_1(\mathbf{D})$ as the largest eigenvalue of $\mathbf{D}$. Since all diagonal entries are finite, one may find $w>0, w \in \mathbb{R}$, such that the matrix $\mathbf{A}=w\mathbf{I}+\mathbf{D} \ge 0$. Since the analyzed network is well connected, matrix $\mathbf{A}$ is irreducible, and accordingly, by the Perron-Frobenius theorem for non-negative irreducible matrices, has one distinctive eigenvalue $\lambda_1(\mathbf{A})$ such that:
\begin{itemize}
    \item $\lambda_1(\mathbf{A})$ is real and positive;
     \item $||\lambda_j(\mathbf{A})|| \le ||\lambda_1(\mathbf{A})||$, $j=\overline{2,nm}$
    \item $\Re\{\lambda_j(\mathbf{A})\} < \lambda_1(\mathbf{A})$, $j=\overline{2,nm}$
\end{itemize}
Since all eigenvectors of $\mathbf{D}$ are simultaneously eigenvectors of $\mathbf{A}$ as well, corresponding eigenvalues are related trough the equation $\lambda_i(\mathbf{A})=w+\lambda_i(\mathbf{D})$. From this equation it is straightforward clear that $\lambda_1(\mathbf{D})$ is real and $\Re\{\lambda_j(\mathbf{D})\} < \lambda_1(\mathbf{D})$, $j=\overline{2,nm}$.
The proof is completed.

From linear systems theory it is a well known that the dynamical system described with the equation (\ref{epsvector}) converges towards the origin, $\mathbf{\epsilon} \rightarrow \mathbf{0}$, as time elapses, provided:
\begin{eqnarray}
Re\{\lambda_{1}(\mathbf{D})\}<0, \nonumber
\end{eqnarray}
or, since $\lambda_{1}(\mathbf{D})$ is real
\begin{eqnarray}
\lambda_{1}(\mathbf{D})<0, \label{trueeigen}
\end{eqnarray}
This implies that all initial perturbations of the dynamical system  (\ref{null}), (\ref{L}), (\ref{H}), and (\ref{transmission})  around the point of epidemic origin (point $\mathbf{0}$) will tend towards the epidemic origin (subsequently the epidemic will vanish from the network), if relation (\ref{trueeigen}) holds. Contrary, for $\lambda_{1}(\mathbf{D})>0$, all initial perturbations will move away from the origin and a state of persistent epidemic presence in the network exists. 

We now show that the following statement holds: Providing relation (\ref{trueeigen}) holds, epidemic origin is globally (on $[0,1]^{n(2^m-1)}$) stable point of equilibrium of the dynamical system 
(\ref{null}), (\ref{L}), (\ref{H}), and (\ref{transmission}). 

The proof will be conducted on the system of relations (\ref{alter1}), a variant that upper-bounds the system (\ref{null}), (\ref{L}), (\ref{H}), and (\ref{transmission}).  In (\ref{alter1}) stochastic independence of joint events is introduced in the term that represent mutation from arbitrary strain $l$ to strain $c$, provided strain $l$ is present and strain $c$ is absent at node $i$ at time $t$. Since, for stochastic events $A$
and $B$, $P(A,B)=P(B \backslash A) P(A) \le  P(A) P(B)$, implementation of stochastic independence of joint events in this case, results in an upper-bound system variant. 

A similar approach is used through-out the literature that studies epidemic spreading on complex networks, primarily in relation to the term defining infection of susceptible node by infected neighbours. The latest, has already been  taken into account in the definition of the system (\ref{null}), (\ref{L}), (\ref{H}), and (\ref{transmission}) as well, for example in the third and fifth term of equation 
(\ref{L}). In what follows, by introducing this notion to one more term, we upper-bound relation (\ref{L}) and even further upper-bound the true behaviour of the system.

Let $p^c_{i}\in [0,1]$ represent the probability that node i is infected with an agent from strain $c$ at an instance of time. Assuming statistical independence of joint events holds, the following system of relations may be considered:
\begin{eqnarray}
\frac{dp^c_{i}}{dt} &=&
(1-p^c_{i})\sum_{l \in H} p^l_{i}  \mu_{l c} + (1-p^c_{i})\sum_{j=1}^n p^c_{j}\sum_{l \in R} A^l_{ij} \beta^c_l - p^c_{i} \left( \gamma_c + \sum_{l \in H } \mu_{c l} \right) = \nonumber\\
&=& (1-p^c_{i})\sum_{l \in H} p^l_{i}  \mu_{l c} +(1-p^c_{i}) \sum_{j=1}^n p^c_{j}t_{i,j}^c  - p^c_{i}\left( \gamma_c + \sum_{l \in H } \mu_{c l} \right) \label{alter1}\\
&\le& \sum_{l \in H} p^l_{i}  \mu_{l c} +\sum_{j=1}^n p^c_{j}t_{i,j}^c  -  p^c_{i}\left( \gamma_c + \sum_{l \in H } \mu_{c l} \right) \nonumber
\end{eqnarray}
The system of relations  (\ref{alter1}) may be re-written in the following form:
\begin{equation} 
    \frac{d\mathbf{p}}{dt} \le \mathbf{D}\mathbf{p}, \label{pvector}
\end{equation}
where $\mathbf{p}=[p_k]$, $k=cn+i$ is a column vector, and $\mathbf{D}$ is an $nm \times nm$ square matrix, as defined above. Consider the system:
\begin{equation} 
    \frac{d\mathbf{p'}}{dt} = \mathbf{D}\mathbf{p'}, \label{p'vector}
\end{equation}
where $\mathbf{p'}$ is an $mn \times 1$ vector. If (\ref{trueeigen}) holds, then the point $\mathbf{p'}=\mathbf{0}$ is globally stable point of equilibrium of the system (\ref{p'vector}). From the  Gronwall-Bellman inequality, there exists a time instance $t_0$, such that for $t>t_0$, $\mathbf{p} \le \mathbf{p'}$ and, since $\mathbf{p} \ge 0$, $\mathbf{p} \rightarrow 0$, when $t \rightarrow \infty$.

Let $L=\{c_{a1},..c_{as}\}$ represent a particular infectious event comprising of $s$ different strain infections. Then:
\begin{eqnarray}
0 \le \lim_{t \rightarrow \infty}p^L_{i}\le \lim_{t \rightarrow \infty}\prod_{c_{r} \in L} p^{c_{r}}_{,i}\prod_{c_r \in H\backslash L} (1-p^{c_{r}}_{i})=0 \times 1=0 \nonumber\\
\end{eqnarray}
The proof is completed. 

Equation (\ref{trueeigen}) defines the epidemic threshold for the dynamical system (\ref{null}), (\ref{L}), (\ref{H}), and (\ref{transmission}).  Though the obtained result is accurate, it gives no straightforward relation between infection parameters of the model and the structure of the network itself. Assessment of the epidemic state of the network requires numerical computation of $\lambda_{1}(\mathbf{D})$, that for spreading processes involving a large number of strains, might be computationally intensive. 

In what follows, we present an analysis, that under some approximations, leads to expressions that roughly estimate the epidemic state of the system as an explicit function of epidemic parameters and the structure of the network. We start by grouping the system of equations (\ref{epsilonapp}) into $n$ groups of $m$ equations and then by summing the equations in each group:
\begin{equation} 
\sum_c \frac{d\epsilon^c_{i}}{dt} =
\sum_{j=1}^n \sum_{l \in R} A^l_{ij} \sum_c\epsilon^c_{j}\beta^c_l  -  \sum_c\epsilon^c_{i} \gamma_c + \sum_c \epsilon^c_{i} \sum_{l \in H^c }\left(\mu_{l c}-\mu_{c l} \right)\label{epsilonsum5}
\end{equation}
Notice that the third term in (\ref{epsilonsum5}), $\sum_c \epsilon^c_{i} \sum_{l \in H^c }\left(\mu_{l c}-\mu_{c l} \right)=0$. Let $\epsilon_i=\sum_c \epsilon^c_{i}$ be the perturbation of the cumulative (all strains considered) infection probability of node $i$ in the vicinity of the epidemic origin. Subsequently, the relation (\ref{epsilonsum5}) may be rewritten in the following form:
\begin{equation} 
\frac{d\epsilon_{i}}{dt} =
\sum_{j=1}^n \sum_{l \in R} A^l_{ij} \sum_c\epsilon^c_{j}\beta^c_l  -  \sum_c\gamma_c \epsilon^c_{i} \label{epsilonsum6}
\end{equation}
One should note that the interpretation of $\epsilon_i$ as a cumulative infection probability is valid only in the vicinity of the origin. Only then one may assume that co-existence of two strains within a single node is highly improbable, and therefor treat infections with different species as mutually exclusive events. Starting from (\ref{epsilonsum6}), from this point onward we consider several scenarios.

\subsection{Bounding}
From (\ref{epsilonsum6}), the following relation may be formulated:
\begin{eqnarray}
\frac{d\epsilon_{i}}{dt} \le \sum_{j=1}^n \epsilon_j\sum_{l \in R} A^l_{ij} \max_c\beta^c_l  -  \min_c \gamma_c \epsilon_i,  \label{epsilonsum6_up}
\end{eqnarray}
where $\beta_l^{\rm sup}=\max_c \beta_l^{c}$, $l=1..R$ and $\gamma^{\rm sup}=\min_c\gamma_c$, are the values of $\beta_l^{c}$, $l=1..R$ and $\gamma_{c}$ that maximize the right-hand side of the equation (\ref{epsilonsum6}) for given $i$. Notice that the choice of these parameters, under the given criteria is independent of $i$. We use this fact to rewrite the relation (\ref{epsilonsum5}) in the following vector form: 

\begin{eqnarray} 
\frac{d\mathbf{\epsilon}}{dt} & \le
\left(\mathbf{T^{\rm sup}} - \gamma^{\rm sup}\mathbf{I}\right)\mathbf{\epsilon},\label{epsilonsup1}
\end{eqnarray}
where $\mathbf{T^{\rm sup}}=[t_{ij}]$, is a $n \times n$ matrix, which elements $t_{ij}=\sum_{l \in R} a^l_{ij} \beta^{\rm sup}_l$ and  $\mathbf{\epsilon}=[\epsilon_1 \; \epsilon_2 \;..\epsilon_N]^T$ is a cumulative (all strains considered) epidemic perturbation vector. Now, consider the system:
\begin{eqnarray} 
\frac{d\mathbf{\epsilon^{\rm sup}}}{dt} &=
\left(\mathbf{T^{\rm sup}} - \gamma^{\rm sup}\mathbf{I}\right) \mathbf{\epsilon^{\rm sup}},\label{epsilonsup2}
\end{eqnarray}
The dynamical system (\ref{epsilonsup2}), is a cumulative variant of the system (\ref{epsilonapp}), and in accordance with the relation (\ref{epsilonsup1}), upper-bounds the behaviour of the dynamical system (\ref{null}), (\ref{L}), (\ref{H}), and (\ref{transmission}) near the region of the epidemic origin (i.e. vector $\mathbf{0}$). From (\ref{epsilonsup2}), it follows that $\mathbf{\epsilon^{\rm sup}} \rightarrow \mathbf{0}$, when $t \rightarrow \infty$, if the larges eigenvalue of the matrix $\mathbf{Q}^{\rm sup}=\mathbf{T^{\rm sup}} - \gamma^{\rm sup}\mathbf{I}$, $\lambda_1(\mathbf{Q}^{\rm sup})$, satisfies the relation $\lambda_1(\mathbf{Q}^{\rm sup})<0$. Since each eigenvector of the matrix $\mathbf{T^{\rm sup}}$ is also an eigenvector of the matrix $\mathbf{Q}^{\rm sup}$, $\lambda_1(\mathbf{Q}^{\rm sup})$, can be expressed as:
\begin{eqnarray}
\lambda_1(\mathbf{Q}^{\rm sup})=\lambda_1(\mathbf{T}^{\rm sup}) - \gamma^{\rm sup}
\end{eqnarray}
In accordance with the derived relations, one may claim that the point $\mathbf{p}=\mathbf{0}$ is a (locally) stable fixed point of the dynamical system (\ref{null}), (\ref{L}), (\ref{H}), and (\ref{transmission}), provided:
\begin{eqnarray}
\lambda_1(\mathbf{T}^{\rm sup})<\gamma^{\rm sup}
\end{eqnarray}
We define:
\begin{eqnarray}
R^+=\frac{\lambda_1(\mathbf{T}^{\rm sup})}{\gamma^{\rm sup}}, \nonumber
\end{eqnarray}
as an upper-bound basic reproductive number. From the discussion above, it is clear that for $R^+<1$ there are no infected nodes. One should notice that, since derivation of $R^+$ was conducted by the analysis of dynamical behaviour of a variant systems that upper-bounds the dynamical behaviour of the system (\ref{null}), (\ref{L}), (\ref{H}), and (\ref{transmission}) near the origin, the inverse does not hold, i.e $R^+>1$ does not necessarily imply that the epidemics exists.

Identically as before, starting from the equation (\ref{epsilonsum6}), and by introducing $\epsilon_i=\sum_c \epsilon^c_{i}$, $\gamma^{\rm inf}=\max_c \gamma_c$, $\beta^{\rm inf}_l=\min_c \beta^c_l$,  $\mathbf{T^{\rm inf}}=[t_{ij}]$, such that $t_{ij}=\sum_{l \in R} A^l_{ij} \beta^{\rm inf}_l$ and $\mathbf{Q}^{\rm inf}=\mathbf{T^{\rm inf}} - \gamma^{\rm inf}\mathbf{I_{N \times N}}$, one obtains:
\begin{eqnarray} \label{epsilonsum4}
\frac{d\mathbf{\epsilon}}{dt} & \ge
\left(\mathbf{T^{\rm inf}} - \gamma^{\rm inf}\mathbf{I_{N \times N}}\right) \mathbf{\epsilon}\nonumber\\ &= \mathbf{Q}^{\rm inf}\mathbf{\epsilon}\nonumber
\end{eqnarray}
The dynamical system:
\begin{eqnarray} \label{epsiloninf}
\frac{d\mathbf{\epsilon^{\rm inf}}}{dt} &=
\left(\mathbf{T^{\rm inf}} - \gamma^{\rm inf}\mathbf{I_{N \times N}}\right) \mathbf{\epsilon^{\rm inf}},\nonumber
\end{eqnarray}
lower-bounds the behaviour of an arbitrary perturbation of the system (\ref{null}), (\ref{L}), (\ref{H}), and (\ref{transmission}) near the epidemic origin, i.e. vector $\mathbf{0}$. Again, the well known  result from linear system analysis suggests that $\mathbf{\epsilon^{\rm inf}} \rightarrow \mathbf{0}$, when $t \rightarrow \infty$, if the larges eigenvalue of the matrix $\mathbf{Q}^{\rm inf}$, satisfies the relation $\lambda_1(\mathbf{Q}^{\rm inf})<0$. By following similar reasoning as before and considering that in this case the analysis of the system (\ref{null}), (\ref{L}), (\ref{H}), and (\ref{transmission})  near the epidemic origin  is conducted by analysis of a lower-bounding variant-system, one may claim that infection in the network  will surely persist, if the following condition holds:
\begin{eqnarray}
\lambda_1(\mathbf{T}^{\rm inf})>\gamma^{\rm inf} 
\end{eqnarray}
By defining the lower-bound basic reproduction number as:
\begin{eqnarray}
R^-=\frac{\lambda_1(\mathbf{T}^{\rm inf})}{\gamma^{\rm inf}}, \nonumber
\end{eqnarray}
one may claim that, for $R^- > 1$ infection surely exists in the observed model described with (\ref{null}), (\ref{L}), (\ref{H}), and (\ref{transmission}).  Contrary, $R^- < 1$ does not guarantee that an ineffective process described with (\ref{null}), (\ref{L}), (\ref{H}), and (\ref{transmission}) will be eradicated from the network as time elapses.
To summarize:
\begin{itemize}
    \item If $R^+<1$ an ineffective process described with the system of equation (\ref{null}), (\ref{L}), (\ref{H}), and (\ref{transmission}) will surely vanish as time elapses.
    \item If $R^->1$ an ineffective process described with the system of equation (\ref{null}), (\ref{L}), (\ref{H}), and (\ref{transmission}) will persist in the network indefinitely.
     \item If $R^+>1$ and $R^-<1$ there is no way to ether confirm nor deny the existence of the epidemic in the network. Only numerical computation of $\lambda_1(\mathbf{D})$ may give a definite answer in this case.
\end{itemize}
In the analysis above, extreme cases scenarios are considered: strains either spread from node to node at maximal spreading rate and die-out with the rate of the most sensitive  strain, or they spread at minimal spreading rate an die out with the rate of the most  endurable strain.

\subsection{Rapid mutation processes}
As a special case, one may consider the system where mutation occurs on a much faster time scale then the infection/die-out process. There are many reasons why such an assumption should hold in practice. First, it is normal to assume that an entity that has capacity to mutate, is designed (biologically or otherwise) to do so several times before it dies out. Next, processes related to small entities (viruses) occur on a much faster time scale then processes related to the host (the node). Concretely, a virus or a bacteria will self replicate and the new entities will undergo a number of biological processes much faster then hosts immune system responds to the epidemic presence or the host is involved in interaction with other entities (nodes) that introduces spreading (these are, finally, the very basic premises on which epidemic spreading models are built). 

For these cases one may assume that once a node is infected, the spreading entity will undergo several mutations before actually spreading to neighboring nodes or dying out. Consider an arbitrary node $i$. Let $\mathbf{\epsilon_i}=[\epsilon_i^1...\epsilon_i^{m}]$. Assuming only mutation occurs, within a reasonable time frame $[0,t_{mut}]$, the dynamical behavior of the infection within node $i$, within the time frame $[0,t_{mut}]$, near the epidemic origin, may be described with the equation:
\begin{eqnarray}
\frac{d\mathbf{\epsilon_i}}{dt} &= \mathbf{M} \mathbf{\epsilon_i}, \label{slow1}
\end{eqnarray}
where $\mathbf{M}=[m_{i,j}]$ is an $R \times R$ matrix, which elements satisfy the relations:
\begin{itemize}
    \item $m_{i,j} = -\sum_c \mu_{ic}$ if $i=j$
    \item $m_{i,j} = \mu_{ji}$ otherwise.
\end{itemize}
It is straightforward obvious that all the columns in $\mathbf{M}$ sum up to zero, therefore $\lambda_1(\mathbf{M})=0$ and $\mathbf{\epsilon}\rightarrow \mathbf{u}=[u_0...u_{m-1}]$, when $t \rightarrow \infty$, where $\mathbf{u}$ is the eigenvector of $\mathbf{M}$ corresponding to $\lambda_1(\mathbf{M})$, and ${\sum_c u_c}=1$.

If now one considers equation (\ref{epsilonsum6}), and assume that the rate of mutation process by far exceeds the rate of spreading/dying (probability of either spreading or dying of a particular specie, within time frame $[0,t_{mut}]$ significantly low), one may claim that the probability of node $i$ being infected with the strain $c$ near the epidemic origin, may roughly be estimated as:
\begin{eqnarray}
\epsilon^c_i \approx u_c \epsilon_i, \nonumber
\end{eqnarray}
which transforms equation (\ref{epsilonsum6}) into:
\begin{eqnarray} 
\frac{d\epsilon_{i}}{dt} &=
\sum_{j=1}^n \epsilon_j \sum_{l \in R} a^l_{ij} \sum_c u_c\beta^c_l  -  \epsilon_i \sum_c\gamma^c u_c \label{epsilonsum7}
\end{eqnarray}
By setting $\beta_l^{\rm eff}=\sum_c u_c\beta^c_l$, $\gamma^{\rm eff}=\sum_c\gamma^c u_c$, $\mathbf{T^{\rm eff}}=[t^{\rm eff}_{i,j}]$, such that $t^{\rm eff}_{i,j}=\sum_{l \in R} a^l_{ij} \beta_l^{\rm eff}$, we can re-write equation (\ref{epsilonsum7}) in the vector form as:
\begin{eqnarray} 
\frac{d\mathbf{\epsilon}}{dt} &=
\left(\mathbf{T^{\rm eff}} - \gamma^{\rm eff}\mathbf{I_{N \times N}}\right) \mathbf{\epsilon},\label{epsilonsum8}
\end{eqnarray}
Using similar arguments as before, we conclude that, provided the rate of mutation exceed the rate of infection/entity extinction, the epidemic threshold of the system (\ref{null}), (\ref{L}), (\ref{H}), and (\ref{transmission}) is defined by: 
\begin{eqnarray}
\lambda_1(\mathbf{T}^{\rm eff})>\gamma^{\rm eff} 
\end{eqnarray}
or in terms of the reproductive number $R^{\rm eff}$:
\begin{eqnarray}
R^{\rm eff}=\frac{\lambda_1(\mathbf{T}^{\rm eff})}{\gamma^{\rm eff}}, \nonumber
\end{eqnarray}

\section{Linear stability analysis of two-contingent models}
Previous section discusses several general properties of our model (\ref{null}), (\ref{L}), (\ref{H}), and (\ref{transmission}). Here we derive close formulae for basic reproduction numbers and epidemic thresholds for various two-contingent models. The disease-free equilibrium is stable if the spectral radius of the Jacobian matrix $\lambda_{\rm max}(J) = \max \{\Re (\lambda) : \lambda \in \text{spectrum of } J \}$ is less than $0$. Hence, the critical values of the transmission rates $(\beta^c_l)$ can be found by examining when $\lambda_{\rm max}(J)$ traverses 0 as $(\beta^c_l)$ are tuned. 
\subsection{Models on multilayer networks}

We  consider the case when the number of contagions is two, while the number of layers is arbitrary. The Jacobian matrix is equal to 
\begin{equation}
J = 
\begin{pmatrix}
T_1 - I(\gamma_1 + \mu_{12}) & I \mu_{21} & T_1 + I \gamma_2 \\
\mu_{12} I & T_2 - I(\gamma_2 + \mu_{21}) & T_2 + I \gamma_1 \\
0 & 0 & -I(\gamma_1 + \gamma_2)  \\
\end{pmatrix}
\label{Jacobian2}
\end{equation}
It is useful to partition the matrix $J$ so that 
\begin{equation*}
J = \begin{pmatrix}
C & D \\
0 & E  \\
\end{pmatrix} 
\end{equation*}
with 
\begin{eqnarray*}
    C &=& 
    \begin{pmatrix}
    T_1 - I(\gamma_1 + \mu_{12}) & I \mu_{21} \\
    \mu_{12} I & T_2 - I(\gamma_2 + \mu_{21})
    \end{pmatrix}
    \nonumber \\
    D &=& 
    \begin{pmatrix}
    T_1 + I\gamma_2 \\
    T_2 + I\gamma_1 
    \end{pmatrix} \\
     E &=& -I(\gamma_1 + \gamma_2)
    \nonumber
\end{eqnarray*}
As we are interested in the case where the maximum eigenvalue of $J$ equals zero, we need to solve $\det J = 0$. The determinant of $J$ takes the form
\begin{equation*}
    \det J = \det C \det E = \det C (-(\gamma_1 + \gamma_2))^n
\end{equation*}
Let $B = T_2 - I(\gamma_2 + \mu_{21})$ and assuming $B$ is invertible, we have
\begin{equation}
    \det C = \det(T_1 - I(\gamma_1 + \mu_{12}) - \mu_{21}IB^{-1}\mu_{12}I) \det B
    \label{BeforeR0}
\end{equation}
Because invertible matrices have non-zero determinants ($\det B \neq 0$), the basic reproduction number is derived from $\det(T_1 - I(\gamma_1 + \mu_{12}) - \mu_{21}IB^{-1}\mu_{12}I) = 0$ to be
\begin{equation} 
    R_0 = \frac{\lambda_{\rm max} (H)}{\gamma_1 + \mu_{12}}. 
    \label{R0}
\end{equation}
where $H = T_1 - \mu_{12} \mu_{21}B^{-1}$. Equation~(\ref{R0}) is a generalization of Eq.~(\ref{eq-rep-number}). Assuming that the transmission rates of contagions 1 and 2 are $\beta^1$
and $\beta^2$, respectively,  on all layers, $R_0 =1 $ in Eq.~(\ref{R0}) defines on the $(\beta^1, \beta^2)$ plane a critical curve, which is, however, not analytically tractable.   
Expressions of the basic reproduction number in terms of eigenvalues of the adjacency matrices, and therefore, critical values of the transmission rates can be derived only for two points on the $(\beta^1, \beta^2)$ plane: $(\beta^1_{ \rm cr}, 0)$ and $(0, \beta^2_{\rm cr})$. 
Indeed, for contagion 1, setting $\beta^2 = 0$ in Equation~(\ref{BeforeR0}) we have
\begin{align}
    \det C & = \det (T_1 - I(\gamma^1 + \mu_{12}) - \mu_{21} I (- I(\gamma_2 + \mu_{21}))^{-1} \mu_{12} I) \nonumber \\
    & = \det (T_1 - I(\gamma_1 + \mu_{12}) + \frac{\mu_{12} \mu_{21}}{\gamma_2 + \mu_{21}} I) \\
    & = \det (T_1 - I(\gamma_1 + \mu_{12} - \frac{\mu_{12} \mu_{21}}{\gamma_2 + \mu_{21}})) 
\end{align}
Similarly for contagion 2, the following reproduction numbers are acquired
\begin{equation}
    R^1_0 = \frac{\lambda_{\rm max}(T_1)}{\left(\gamma_1 + \mu_{12} - \frac{\mu_{12} \mu_{21}}{\gamma_2 + \mu_{21}} \right)}; \quad R^2_0 = \frac{\lambda_{\rm max}(T_2)}{\left(\gamma_2 + \mu_{21} - \frac{\mu_{12} \mu_{21}}{\gamma_1 + \mu_{12}} \right)}
\end{equation}
The critical transmission rates can be derived by assuming that $\lambda_{\rm max}(T_c) = \beta^c \lambda_{\rm max} (\sum_{l \in R} A_l)$ and  $R^1_0 = R^2_0 = 1$. Therefore, 
\begin{align}
    \beta^1_{\rm cr} = \frac{1}{\lambda_{\rm max}(\sum_{l \in R} A_l)} \left(\gamma_1 + \mu_{12} - \frac{\mu_{12} \mu_{21}}{\gamma_2 + \mu_{21}} \right), \label{beta_critical-1} \\
    \beta^2_{\rm cr} = \frac{1}{\lambda_{\rm max}(\sum_{l \in R} A_l)} \left(\gamma_2 + \mu_{21} - \frac{\mu_{12} \mu_{21}}{\gamma_1 + \mu_{12}} \right). \label{beta-critical-2}
\end{align}
This is a generalization of (\ref{eq-beta-cr}). For the single layer case and $\mu_{12} = 0$ it can be seen that (\ref{beta_critical-1}) reduces to the well known critical value $\beta_{\rm cr} = \gamma / \lambda_{\rm max}(A)$.

\begin{figure}
    \centering
    \includegraphics[scale=0.70]{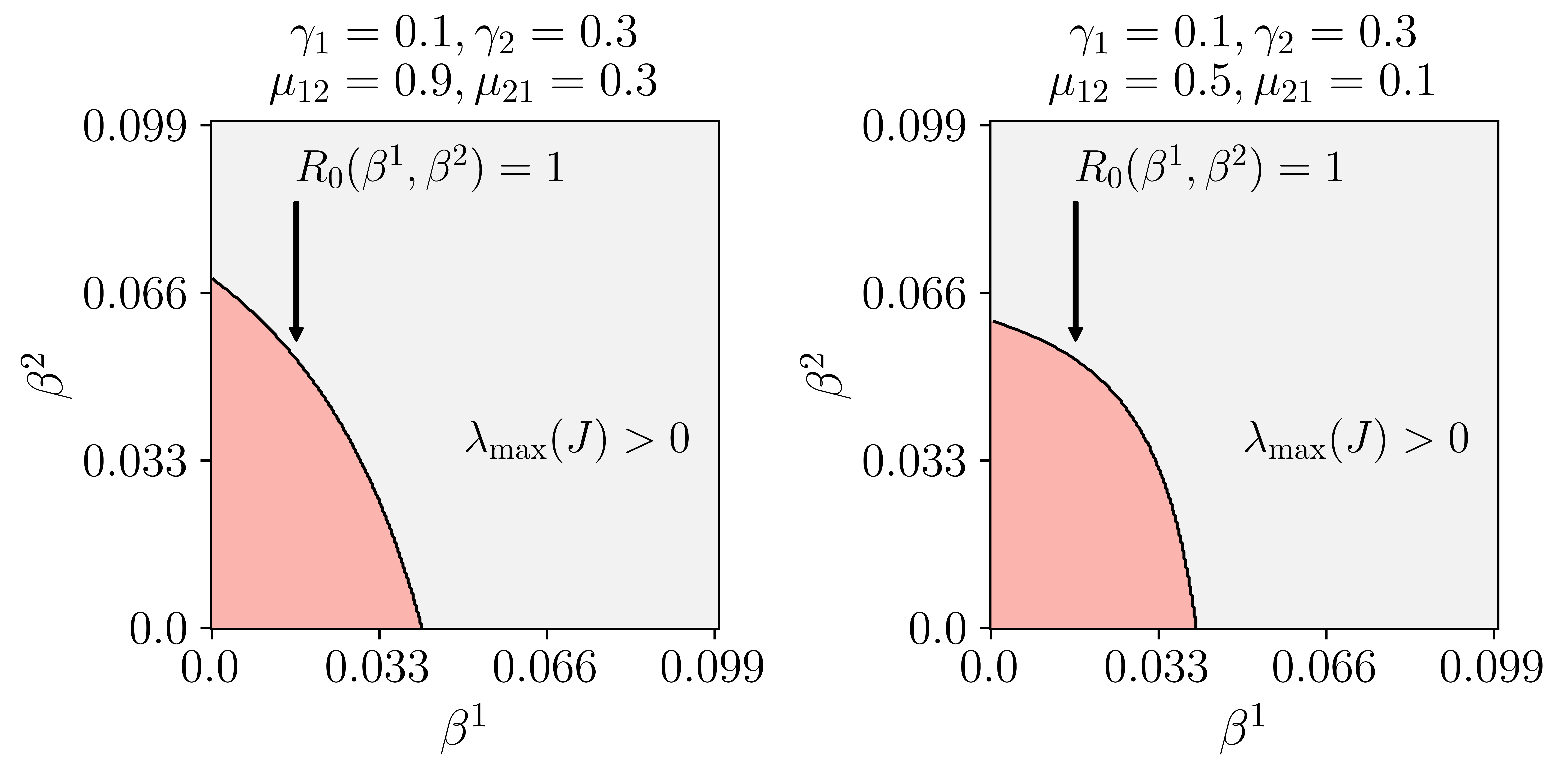}
    \caption{Sign of the spectral radius of $J$ as a function of the transmission rates $\beta^1$ and $\beta^2$ for different recovery and mutation rates. The numerical simulations are conducted on a two-layer network with $n = 50$ nodes where the first layer is an Erdos-Renyi network with parameter $p = 0.05$, and the second layer is a Barabasi-Albert network with parameter $k = 2$. The basic reproduction number is computed using equation (\ref{R0}) for every pair $(\beta^1, \beta^2)$ and the curve where $R_0 = 1$ is drawn. }
    \label{fig:Figure4}
\end{figure}

Now, we provide numerical simulations on random networks to investigate the correctness of the derived basic reproduction numbers and the critical transmission rates.  
The numerical simulations are conducted on a two-layer network with $n = 50$ nodes where the first layer is an Erdos-Renyi network with parameter $p = 0.05$, and the second layer is a Barabasi-Albert network with parameter $k = 2$. Figure \ref{fig:Figure4} depicts the sign of the spectral radius of the Jacobian matrix (\ref{Jacobian2}) as the transmission rates $\beta^1$ and $\beta^2$ are ranged from $0$ to $0.1$. For every pair $(\beta^1, \beta^2)$ the basic reproduction number $R_0$ is computed using Equation (\ref{R0}) and the curve where $R_0 = 1$ is drawn. It can be seen that for $R_0 < 1$ the disease-free equilibrium stable and for $R_0 > 1$ it is not. The critical values, Eq.~(\ref{beta_critical-1}) and (\ref{beta-critical-2}),
are computed to be $\beta^1_{\rm cr } = 0.041$ and $\beta^2_{\rm cr} = 0.069$ for $\gamma_1 = 0.1, \gamma_2 = 0.3, \mu_{12} = 0.9, \mu_{21} = 0.3$, and $\beta^1_{\rm cr } = 0.04$ and $\beta^2_{\rm cr} = 0.061$ for $\gamma_1 = 0.1, \gamma_2 = 0.3, \mu_{12} = 0.5, \mu_{21} = 0.3$.

\subsection{Competition models on multilayer networks}

We now consider the case of our coinfection model when each node can either be susceptible or infected by only a single disease, yielding a competition model. We further assume that each disease can be transmitted through one layer (disease c transmits on layer c), so that $\beta^c \equiv \beta^c_l \neq 0$ only when $c = l$. 
Since in this case $|L| \leq 1$, the model equations (4) become
\begin{equation}
    \frac{d p^c_i}{dt} = \sum_{l=1}^L p^l_i \mu_{lc} + \left(1 - \sum_{l = 1}^L p^l_i\right) \sum_{j=1}^n A^c_{ij} \beta^c p^c_j - p^c_i \left(\gamma_c + \sum_{l = 1}^L \mu_{cl}\right).
\end{equation}
for $c = 1, \ldots, m$ and $i = 1, \ldots, n$. Similarly to the previous section we are interested in the stability of the disease-free equilibrium of the competition model. The Jacobian matrix takes the form
\begin{equation}
    J = \bigoplus_{c = 1}^L \left( \beta^c A_c - \left(\gamma_c + \sum_{l=1}^L \mu_{cl}\right) I_{n \times n}\right) + \Gamma \otimes I_{n \times n}. 
\end{equation}
where $\bigoplus$ denotes the direct sum of $L$ matrices and $\Gamma$ is the mutation rate matrix
\begin{equation}
    \Gamma = 
    \begin{pmatrix}
    0 & \mu_{21} & \dots & \mu_{L1} \\
    \vdots & \vdots & \ddots & \vdots \\
    \mu_{1L} & \mu_{12} & \dots & 0
    \end{pmatrix}.
\end{equation}
For the two-contingent case the linear stability analysis of the disease-free equilibrium is similar to that of the coinfection model with two-contingents. The Jacobian takes the form
\begin{equation}
    J = \begin{pmatrix}
    \beta^1 A_1 - I(\gamma_1 + \mu_{12}) & I \mu_{21} \\
    \mu_{12} I & \beta^2 A_2 - I(\gamma_2 + \mu_{21})
    \end{pmatrix}
\end{equation}
the basic reproduction number becomes
\begin{equation}
    R_0 = \frac{\lambda_{\rm max} (\beta^1 A_1 - \mu_{12}\mu_{21} (\beta^2 A_2 - (\gamma_2 + \mu_{21})I)^{-1})}{\gamma_1 + \mu_{12}}
    \label{R0Comp}
\end{equation}
and the critical transmission rates have the following simplified form
\begin{equation}
    \beta^1_{\rm cr} = \frac{\gamma_1 + \mu_{12} - \frac{\mu_{12} \mu_{21}}{\gamma_2 + \mu_{21}}}{\lambda_{\rm max}(A_1)}; \quad
    \beta^2_{\rm cr} = \frac{\gamma_2 + \mu_{21} - \frac{\mu_{12} \mu_{21}}{\gamma_1 + \mu_{12}}}{\lambda_{\rm max}(A_2)}.
    \label{beta_criticalcomp}
\end{equation}

\begin{figure}
    \centering
    \includegraphics[scale=0.7]{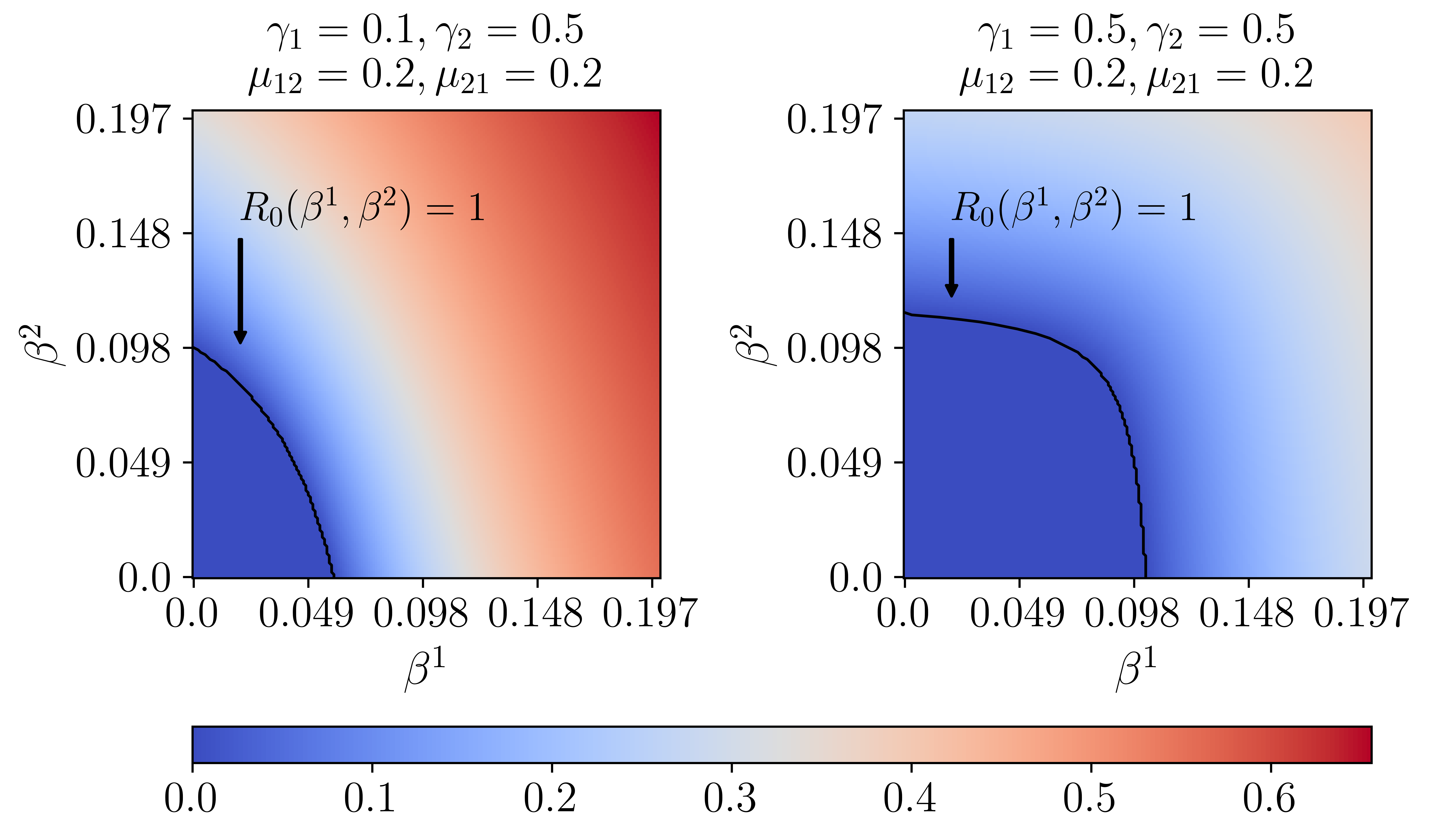}
    \caption{Fraction of infected nodes as a function of the transmission rates $\beta^1$  and $\beta^2$ for different recovery and mutation rates. The numerical simulations are conducted on two-layer Erdos-Renyi random networks with $n = 50$ nodes and $p = 0.1$. The basic reproduction number is computed using equation (\ref{R0Comp}) for every pair $(\beta^1, \beta^2)$ and the curve where $R_0 = 1$ is drawn.}
    \label{fig:Figure5}
\end{figure}

We conduct numerical simulations on two-layer Erdos-Renyi random networks with $n = 50$ nodes and edge probability $p = 0.1$. Figure \ref{fig:Figure5} depicts the fraction of infected nodes at the steady state, as the transmission rates $\beta^1$ and $\beta^2$ are ranged from 0 to 0.2. For every pair $(\beta^1, \beta^2)$ the basic reproduction number $R_0$ is computed using Equation (\ref{R0Comp}) and the curve where $R_0 = 1$ is drawn. It can be seen that for $R_0 < 1$ the fraction of infected nodes is 0, and for $R_0 > 1$, $\sum_{i=1}^n p^1_i + p^2_i > 0$. The critical values, Eq.~(\ref{beta_criticalcomp}),
are computed to be $\beta^1_{\rm cr } = 0.06$ and $\beta^2_{\rm cr} = 0.098$ for $\gamma_1 = 0.1, \gamma_2 = 0.3, \mu_{12} = 0.2, \mu_{21} = 0.2$, and $\beta^1_{\rm cr } = 0.104$ and $\beta^2_{\rm cr} = 0.114$ for $\gamma_1 = 0.5, \gamma_2 = 0.5, \mu_{12} = 0.2, \mu_{21} = 0.2$.

\subsection{Compartmental models}

Multiple-contagion compartmental model consists of $2^m$ compartments. Dynamical equations describing the time evolution of the compartmental model are acquired by removing variables that are network dependent. As an example, we consider 2 contingents (resulting in 4 compartments), $d$ layers,  and write $p^\emptyset = p_0$, $p^{\{1\}} = p_1$, $p^{\{2\}} = p_2$, and $p^{\{1,2\}} = p_3$. Then the model equations can be written as:
\begin{align*}
\dot{p}_0 & =  p_1\gamma_1 + p_2 \gamma_2 - p_0 \left( (p_1 + p_3) \sum_{c=1}^d \beta^1_c + (p_2 + p_3) \sum_{c=1}^d \beta_c^2  \right)  \\
\dot{p}_1 & = p_2\mu_{21} +p_3\gamma_2 + p_0 \left( (p_1 + p_3) \sum_{c=1}^d \beta^1_c \right)
 - p_1 \left( \gamma_1 +  \mu_{12} + (p_2 + p_3) \sum_{c=1}^d \beta_c^2 \right) \\
\dot{p}_2 & =  p_1\mu_{12} +p_3\gamma_1 + p_0 \left( (p_2 + p_3) \sum_{c=1}^d \beta^2_c \right)  - p_2 \left( \gamma_2 +  \mu_{21} + (p_1 + p_3) \sum_{c=1}^d \beta_c^1 \right) \\
\dot{p}_3 & =  p_2 (p_1 + p_3) \sum_{c=1}^d \beta_c^1 + p_1 (p_2 + p_3) \sum_{c=1}^d \beta_c^2  - p_3 \left( \gamma_1 + \gamma_2   \right)
\end{align*}
Since $p_0(t) +p_1(t) +p_2(t) +p_3(t) = 1$ for all $t$, we keep only $p_1,p_2,p_3$ variables. The  Jacobian matrix of the model evaluated at the disease-free equilibrium is a $3\times 3 $ matrix: 
\begin{equation}
J = 
\begin{pmatrix}
    \sum_{l = 1}^d \beta^{1}_l - (\gamma^1 + \mu^{12}) & \mu^{21} & \sum_{l = 1}^d \beta^{1}_l + \gamma^2 \\
    \mu^{12} & \sum_{l = 1}^d \beta^{2}_l - (\gamma^2 + \mu^{21}) & \sum_{l = 1}^d \beta^{2}_l + \gamma^1 \\
    0 & 0 & -(\gamma^1 + \gamma^2)
\end{pmatrix}.
\label{JacCompartmental}
\end{equation}
One eigenvalue is equal to $-(\gamma^1 + \gamma^2)$ and hence is always negative. The other two eigenvalues of the model are the eigenvalues of the matrix $A$ defined as: 
\[
A = 
\begin{pmatrix}
    \sum_{l = 1}^d \beta^{1}_l - (\gamma^2 + \mu^{21}) & \mu^{21} \\
    \mu^{12} & \sum_{l = 1}^d \beta^{2}_l - (\gamma^2 + \mu^{21})  
\end{pmatrix}
\]
The largest eigenvalue of the Jacobian $J$ is, therefore, given by:  
$$
\lambda_{\rm max} = \frac{a + b + \sqrt{(a + b)^2 - 4(ab - \mu^{12}\mu^{21})}}{2}
$$
where $a = \sum_{l = 1}^d \beta^{1}_l - (\gamma^1 + \mu^{12})$ and $b = \sum_{l = 1}^d \beta^{2}_l - (\gamma^2 + \mu^{21})$. The last equation can be rewritten as  
$$
\lambda_{\rm max} = 
\frac{a + b + \sqrt{(a - b)^2 + 4 \mu^{12}\mu^{21}}}{2} = 
\begin{cases}
a + \epsilon & a \geq b \\
b + \epsilon & a < b
\end{cases},
$$
where $\epsilon$ is a function of $\mu^{12} $ and $\mu^{21}$.
Setting $\lambda_{\rm max} = 0$, the basic reproduction number becomes:  
$$
R_0 = 
\begin{cases}
\dfrac{\sum_{l = 1}^d \beta^1_l + \epsilon}{\gamma^1 + \mu^{12}} & a \geq b \\
\dfrac{\sum_{l = 1}^d \beta^2_l + \epsilon}{\gamma^2 + \mu^{21}} & a < b .
\end{cases}
$$

\begin{figure}
\begin{center}
\includegraphics[scale=0.70]{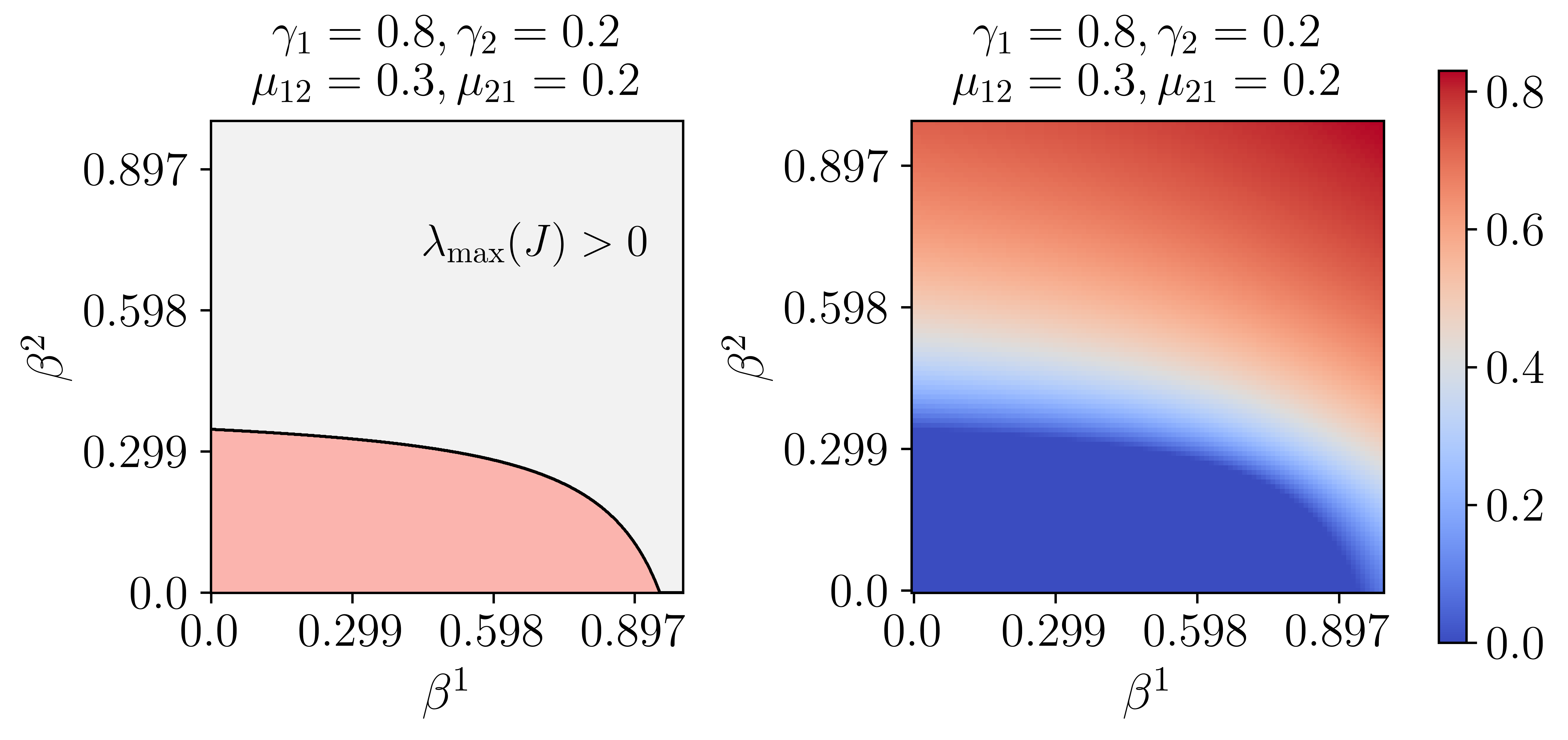}
\end{center}
\caption{Left: The sign of the largest eigenvalue of the Jacobian matrix Eq. (\ref{JacCompartmental}) as a function of $\beta^1$ and $\beta^2$; the rest of the parameters are fixed to $\gamma_1 = 0.8$, $\gamma_2 = 0.2$, $\mu_{12} = 0.3$, $\mu_{21} = 0.2$. The curve indicates $\beta^2_{cr}$ as a functions of $\beta^1_{cr}$ given by Eq.~(\ref{eq-beta2-vs-beta1}). Right: Fraction of infected nodes as a function of the transmission rates $\beta^1$  and $\beta^2$.} \label{fig-2}
\end{figure}

\begin{figure}
    \centering
    \includegraphics[scale=0.75]{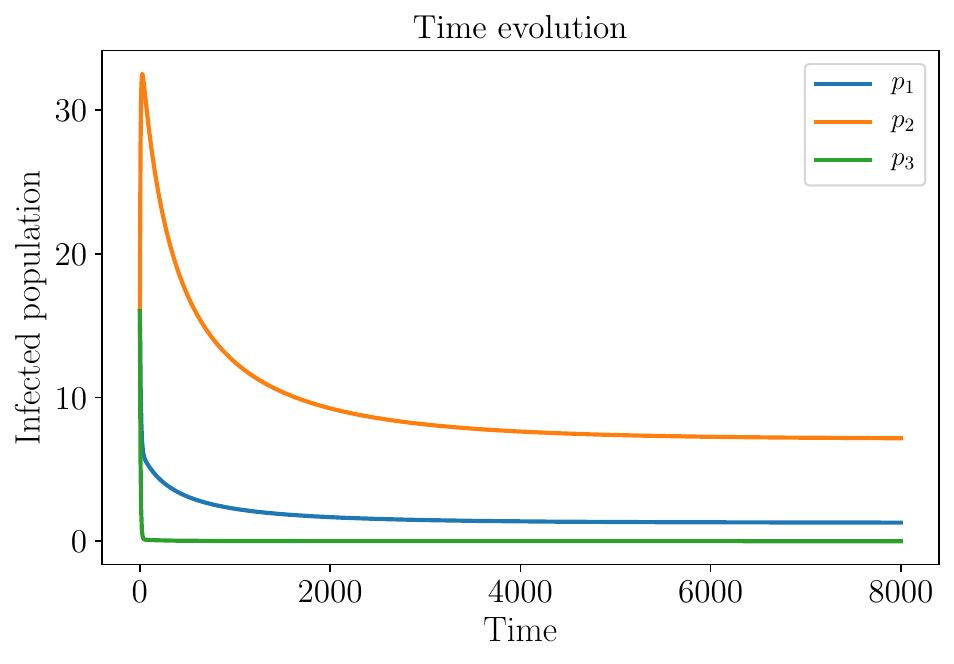}
    \caption{Size of the infected population as a function of time for $\beta^1 = 0$ and $\beta^2 = 0.352$. The parameter $\beta^2$ is chosen such that is slightly larger than the critical value $\beta^2_{cr} = 0.345$.}
    \label{fig:fig7}
\end{figure}

We consider two examples. In the first example we assume $d=1$ and let  $\lambda_{\rm max}=0$ (equivalently $R_0=1$). Then,  the critical values of $\beta_1^1$ and $\beta_2^2$, $\beta^1_{cr}$ and $\beta^2_{cr}$, are related to each other through equation 
$ ab = \mu^{12} \mu^{21}  $, which can be rewritten as  
\begin{equation} 
\beta_{\rm cr}^2 = \dfrac{\mu_{12}\mu_{21}}{\beta^1_{\rm cr} - \gamma_1 - \mu_{12}} + \gamma_2 + \mu_{21}
\label{eq-beta2-vs-beta1}
\end{equation}
for $ 0 \leq \beta^1_{cr} \leq c_1$ where $c_1>0$ is a constant that depends on the parameters $\mu^{12}$, $\mu^{21}$, $\gamma^1$, and $\gamma^2$. Assume now that $\gamma^1 = 0$, then for $\beta^1_{cr}=0$ and $\beta^2_{cr} =\gamma^2$, $\lambda_{\rm max}=0$, while for $\beta^1=0$ and $\beta^2 > \gamma^2$, $\lambda_{\rm max}>0$. Therefore, even when $\beta^1 =0$, the epidemic state is present in the model (when $\beta^2 > \gamma^2$). This is illustrated in Figure \ref{fig:fig7}. We fix the parameters to $\gamma^1 = 0.8$, $\gamma^2 = 0.2$, $\mu^{12} = 0.3$, $\mu^{21} = 0.2$, 
and plot $\beta_{cr}^2$ vs $\beta^1_{cr}$ on Figure \ref{fig-2}. For $\beta^1_{cr}=0$, we have $\beta_{cr}^2=0.345$. Figure \ref{fig:fig7} shows how the size of the infected population evolves over time for $\beta^1 = 0$ and $\beta^2 = 0.352$: from a population of 400 approximately 7.142 of the population are infected with infection 2, 1.293 with infection 1, 0.005 with both infections 1 and 2, and 391.56 are susceptible. For any $\beta^2 > \beta^2_{cr}$, the model approaches a fixed point solution different than the origin, although the values of $p^1$, $p^2$, and $p^{12}$ are small (of order of  $\beta^2 - \beta^2_{cr}$).   

For the second example, we assume that $d=2$, and write $\beta = \beta^1_1 = \beta^2_2$, and $\hat{\beta} = \beta^1_2 = \beta^2_1$.  Setting  $\lambda_{\rm max}=0$ (equivalently $R_0=1$) leads to the following two relations between the critical values of $\beta$ and $\hat{\beta}$, $\beta_{cr}$ and $\hat{\beta}_{cr}$,  
\begin{eqnarray*}
(\hat{\beta}_{cr})^2 + \hat{\beta}_{cr} (2 \beta_{cr} + c + d) + (\beta_{cr} + c)(\beta_{cr} + d) - \mu^{12}\mu^{21} &=& 0 \\
({\beta}_{cr})^2 + {\beta}_{cr} (2 \hat{\beta}_{cr} + c + d) + (\hat{\beta}_{cr} + c)(\hat{\beta}_{cr} + d) - \mu^{12}\mu^{21} &=& 0 
\end{eqnarray*}
which, after some algebra, can be written as $\hat{\beta}_{cr} = \beta_{cr} + c$, 
for $ 0 \leq \beta^1_{cr} \leq c_2$ where $c_2>0$ and $c>0$ are constants that depend on the parameters $\mu^{12}$, $\mu^{21}$, $\gamma^1$, and $\gamma^2$. Again the absence of epidemic threshold ($\beta =0$) is observed while the epidemic state is presented (when $\hat{\beta}>c$).

Finally, we remark that in a special case when $\beta_l^1 = \beta_l^2 \equiv \beta_l $ for all $l$ and $\gamma_1 = \gamma_2 \equiv \gamma$, the basic reproduction number reduces to:
$$
R_0 = \frac{\sum_l \beta_l}{\gamma}.
$$
for arbitrary values of $\mu^{12}$ and $\mu^{21}$. Therefore, in this example, mutation does not influence the basic reproduction number and the corresponding threshold as long as the transmission and recovery rates of two contingents are equal to each other. Moreover, if, in addition, $\beta_l = \beta$ for all $l$, the basic reproduction number reduces to 
$$
R_0 = \frac{d\beta}{\gamma}.
$$

\section{Conclusion}
In summary, we have suggested a multiple contagion SIS model on a multilayer network that incorporates different spreading channels and disease mutations. The model is analytically tractable;  in particular, the absence of epidemic threshold and critical behavior even for compartmental models could further improve our understanding of epidemic spreading. The results obtained here could have implications in various disciplines including epidemiology, biology, and sociology.

\section*{Acknowledgment}

This research was supported in part by DFG (grant \textit{Random search processes, L\'{e}vy flights, and random walks on complex networks}) and ONR/ONR Global (grant No. N62909-16-1-2222).

\section*{References}
\bibliographystyle{model1-num-names}
\bibliography{sample.bib}
\end{document}